\documentclass{iaus}
\usepackage{graphicx}

\title[IAUS 249.~~SPH simulations of star/planet formation] 
{SPH simulations of star/planet formation triggered by cloud-cloud collisions}

\author[S. Kitsionas, A. Whitworth \& R. Klessen]   
{Spyridon Kitsionas$^1$,
Anthony P. Whitworth$^2$
\and Ralf S. Klessen$^3$}

\affiliation{$^1$Astrophysikalisches Institut Potsdam, An der Sternwarte 16, D-14482 Potsdam, Germany \\ email: {\tt skitsionas@aip.de} \\ [\affilskip]
$^2$School of Physics \& Astronomy, Cardiff University, P.O. Box 913, CF24 3AA Cardiff, U.K. \\ [\affilskip]
$^3$Institut f\"ur Theoretische Astrophysik, Universit\"at Heidelberg, \\ Albert-Ueberle-Str. 2, D-69120 Heidelberg, Germany}

\pubyear{2008}
\volume{249}  
\pagerange{1--6}
 \jname{Exoplanets: Detection, Formation and
Dynamics} \editors{Y.-S. Sun, S. Ferraz-Mello \& J.-L. Zhou, eds.}
\begin{document}

\maketitle

\begin{abstract}
We present results of hydrodynamic simulations of star formation triggered by 
cloud-cloud collisions. During the early stages of star formation, low-mass 
objects form by gravitational instabilities in protostellar discs. A number of 
these low-mass objects are in the sub-stellar mass range, including a few 
objects of planetary mass. The disc instabilities that lead to the 
formation of low-mass objects in our simulations are the product of 
disc-disc interactions and/or interactions between the discs and their 
surrounding gas.
\end{abstract}

\firstsection 
\section{Introduction}

In the classical picture (\cite[Shu, Adams \& Lizano 1987]{shu_etal87}) stars 
form from the collapse of a single gaseous core in isolation: all stages 
of the star formation process (i.e. objects of Class 0, I, II, III) as well as 
planet formation are terminated before the newly formed star/planetary system 
interacts with its environment (e.g. other stars/planetary systems, and/or 
its surrounding gas). In the more dynamic star formation paradigm of 
gravoturbulent fragmentation 
(\cite[Mac~Low \& Klessen 2004]{maclow_klessen04}) interactions during the 
formation process are not only a common phenomenon but also a sufficient 
condition for the formation of binary and higher-order multiple systems, as 
well as for explaining the shape of the Initial Mass Function
(\cite[Jappsen \etal\ 2005]{jappsen_etal05}; see also 
\cite[Bate, Bonnell \& Bromm 2003]{bate_etal03}, \cite[Bate \& Bonnell 2005]{bate_bonnell05}), 
the evolution of the angular momentum distribution of protostars 
(\cite[Jappsen \& Klessen 2004]{jappsen_klessen04}), their mass accretion 
rates (\cite[Schmeja \& Klessen 2004]{schmeja_klessen04}), the 
spatial distribution of young stellar objects 
(\cite[Schmeja \& Klessen 2006]{schmeja_klessen06}), etc. The competing 
paradigm of a more isolated star formation process advocated by 
\cite[Krumholz \& McKee (2005)]{krumholz_mckee05}, uses large-scale turbulent 
support against collapse in a rather isotropic way, giving rise to 
supercritical cores with a variety of masses (see e.g.
\cite[Krumholz, McKee \& Klein 2005]{krumholz_etal05} {\it vs}.
\cite[Bonnel \& Bate 2006]{bonnel_bate06}). In contrast, gravoturbulent star 
formation is based on the small-scale density fluctuations produced by 
supersonic turbulence, with gravity then promoting the strongest of these 
fluctuations, determining the onset of collapse.

In the context of gravoturbulent fragmentation, 
\cite[Chapman \etal\ (1992)]{chapman_etal92}, 
\cite[Turner \etal\ (1995)]{turner_etal05}, 
\cite[Whitworth \etal\ (1995)]{whitworth_etal95}, 
and \cite[Bhattal \etal\ (1998)]{bhattal_etal98} have studied star formation 
triggered by cloud-cloud collisions. In essence, the cloud-cloud collision 
mechanism describes the small-scale evolution of the interaction between 
colliding flows (\cite[V\'azquez-Semadeni \etal\ 2006, 2007]{vazquez_etal06}). 
We note that several authors have studied the interaction between 
large-scale colliding flows in non-self-gravitating media, discussing the 
importance of such interactions for the formation of molecular-cloud structure 
and turbulence (e.g. \cite[Hunter \etal\ 1986]{hunter_etal06}, 
\cite[Walder \& Folini 1998]{walder_folini98}, 
\cite[Heitsch \etal\ 2005]{heitsch_etal05}, 
\cite[Hennebelle \& Audit 2007]{hennebelle_audit07}). These authors discuss 
the formation at best of low-mass structures such as dense cloud cores, 
whereas in our self-gravitating simulations we use such dense cores as the 
initial conditions in order to follow star formation.

In this paper we present the results of recent high-resolution hydrodynamic 
simulations of collisions between low-mass clumps 
(\cite[Kitsionas \& Whitworth 2007]{kitsionas_whitworth07}). Due to the 
high numerical resolution of our simulations, we can resolve, 
for the first time in simulations of cloud-cloud collisions, the formation of 
sub-stellar objects, some of which have a mass that lies close to the boundary 
between brown dwarves and planetary-mass objects, i.e. $\sim 15 M_J$. These 
planetary-mass objects form by gravitational instabilities in the disc of 
their parent protostar, i.e. by the same mechanism that is responsible for the 
formation of low-mass stellar and/or brown dwarf companions to the central 
protostar of such discs. Moreover, the disc instabilities that lead to the 
formation of low-mass objects in our simulations are the product of 
disc-disc interactions and/or interactions between the discs and their 
surrounding gas.
Furthermore, we deal here with disc instabilities of young, massive discs 
still in the process of formation, i.e. at the early stages of the disc 
life. We therefore caution the reader not to confuse the instabilities 
reported here with the extensive work on gravitational instabilities and 
related phenomena in thin, or in general evolved discs that have been recently 
reviewed by \cite[Durisen \etal\ (2007)]{durisen_etal07}. As for the core 
accretion model of planet formation, we refer the reader to a number of papers 
in this volume dealing with the different stages of solid core formation, gas 
accretion and migration in protoplanetary discs.

In section 2, we give a brief description of the model and the numerical 
method we use. In section 3, we present our results. In section 4, we discuss 
our findings with respect to models of gas giant planet formation due to 
gravitational instabilities in protostellar discs.

\section{Our model and numerical method}

In \cite[Kitsionas \& Whitworth (2007)]{kitsionas_whitworth07}, we 
investigate, by means of numerical simulations, the phenomenology 
of star formation triggered by low-velocity collisions between low-mass 
molecular clumps. The simulations are performed using a smoothed particle 
hydrodynamics (SPH) code which 
satisfies the Jeans condition by invoking on-the-fly Particle Splitting 
(\cite[Kitsionas \& Whitworth 2002]{kitsionas_whitworth02}).

Clumps are modelled as stable truncated (non-singular) isothermal, i.e. 
Bonnor-Ebert, spheres. 
Collisions are characterised by $M_0$ (clump mass), $b$ (offset parameter, 
i.e. ratio of impact parameter to clump radius), and ${\cal M}$ (Mach Number, 
i.e. ratio of collision velocity to effective post-shock sound speed). 
The gas follows a barotropic equation of state, which is intended to 
capture (i) the scaling of pre-collision internal velocity dispersion with 
clump mass (\cite[Larson 1981]{larson81}), (ii) post-shock 
radiative cooling, and (iii) adiabatic heating in optically thick 
protostellar fragments (with exponent $\gamma \simeq 5/3$). The equation of 
state we use is given by Eq. 1 of 
\cite[Kitsionas \& Whitworth (2007)]{kitsionas_whitworth07} and is graphically 
illustrated in the temperature-density plane by their Fig. 1. In short, the 
initial effective isothermal sound speed of the gas, i.e. when non-thermal 
pressure due to turbulence is included, follows the 
\cite[Larson (1981)]{larson81} relations in order to model the observed 
variation of the internal velocity dispersion with clump mass. As long as the 
collisions begin, the 
effective sound speed is reduced according to a $P \propto \rho^{1/3}$ 
relation until it reaches a value equivalent to the temperature of 10 K. Then 
the gas remains isothermal with increasing density up to the point where it 
becomes opaque to its own cooling radiation and heats up as 
$P \propto \rho^{5/3}$.

\begin{figure}[t]
\begin{center}
\resizebox{6.5cm}{!}{\includegraphics{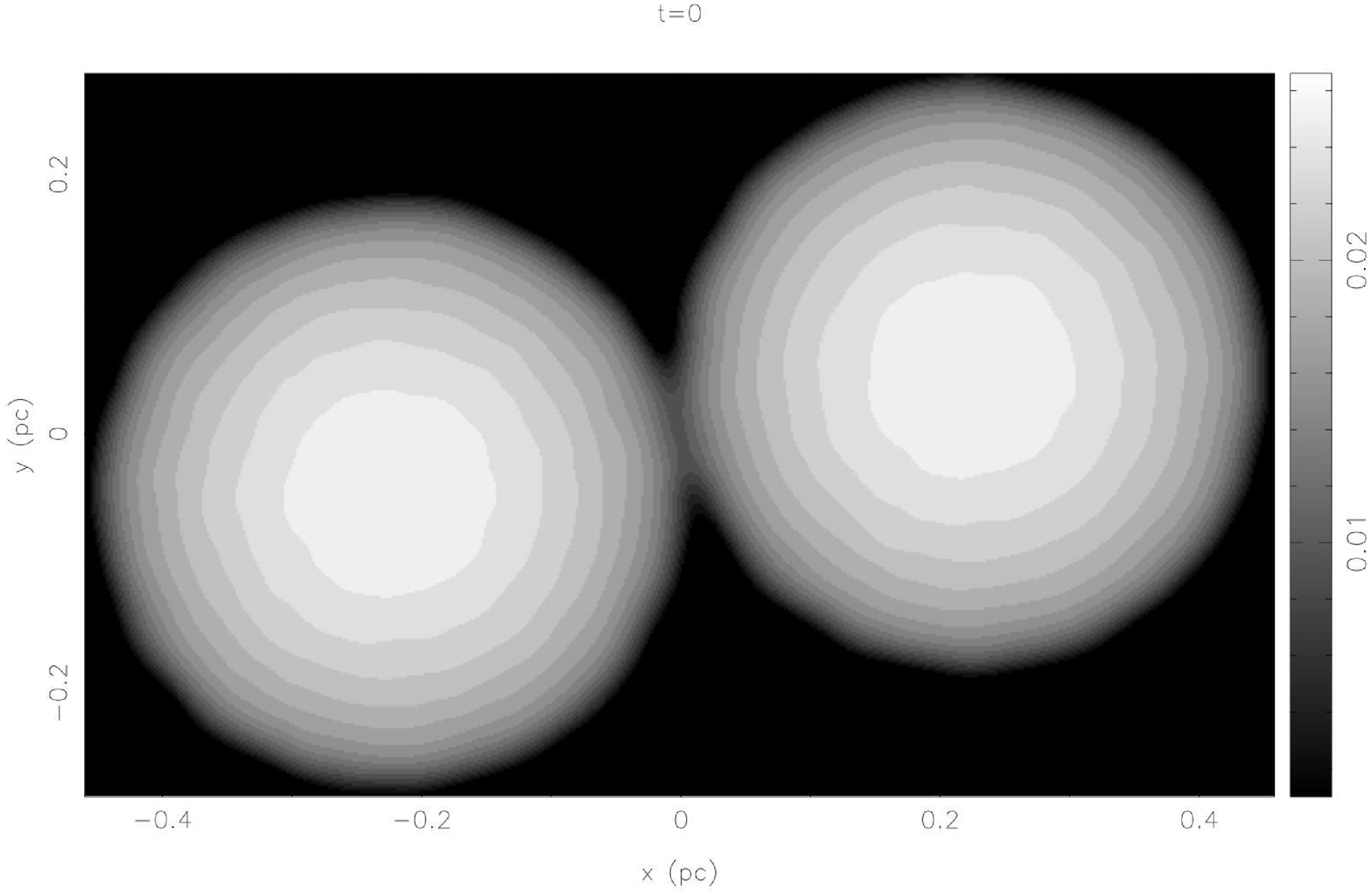} }
\resizebox{6.5cm}{!}{\includegraphics{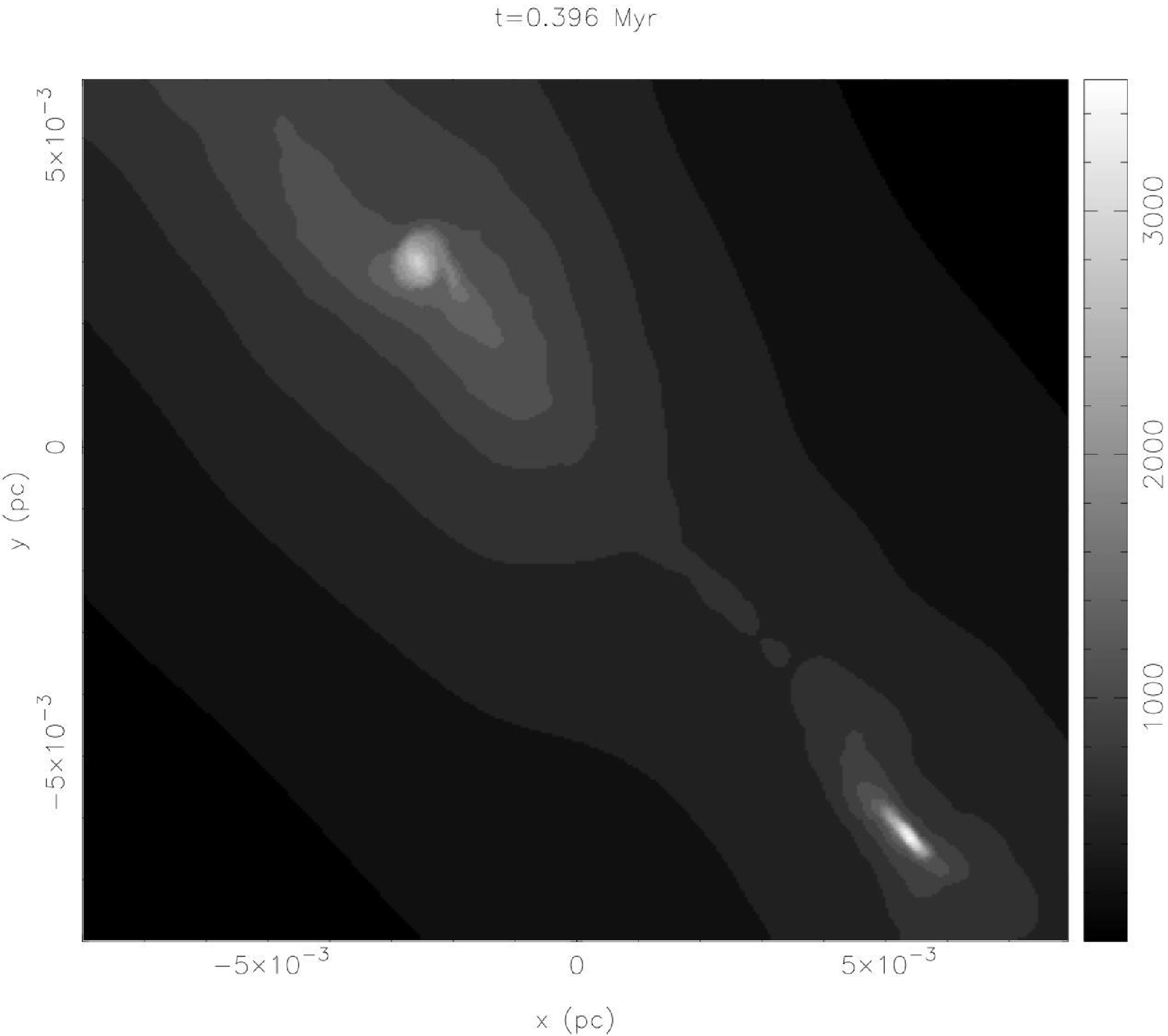} }
\caption{Column density plots for the collision with $M_0 = 10 M_\odot$, 
$b=0.2$, ${\cal M}=10$. 
{\it Left Panel.} Initial conditions viewed along the 
$z$-axis; $\Delta x = 0.92\,{\rm pc}, \; \Delta y = 0.56\,{\rm pc}$; the 
grey-scale is logarithmic, in units of g cm$^{-2}$, with sixteen equal 
intervals from $1.00 \times 10^{-3} \, {\rm g} \, {\rm cm}^{-2}$ to 
$2.69 \times 10^{-2} \, {\rm g} \, {\rm cm}^{-2}$. 
{\it Right Panel.} $t = 0.396\,{\rm Myr}$; viewed 
along the $z$-axis; $\Delta x = 0.016\,{\rm pc}, \; \Delta y = 0.014\,{\rm pc}$; sixteen-interval logarithmic grey-scale, in units of g cm$^{-2}$, from 
$2.14 \times 10^{-1} \, {\rm g} \, {\rm cm}^{-2}$ to 
$3.55 \times 10^{3} \, {\rm g} \, {\rm cm}^{-2}$.}
   \label{fig1}
\end{center}
\end{figure}

From the variety of models we have investigated, we present here the collision
between two $M_0 = 10 M_\odot$ clumps, each moving with velocity (along the 
$x$-axis) $v_{\rm clump} = 1$ km s$^{-1}$ (corresponding to a Mach number 
${\cal M}=10$, i.e. assuming a post-shock sound speed of 0.2 km s$^{-1}$), and 
with an offset parameter $b=0.2$ taken along the $y$-axis. The initial 
conditions are shown in the left panel of Fig. \ref{fig1}.

We evolve the simulations using an SPH code. This is a Lagrangian method for 
3D hydrodynamics. In SPH the sampling points (particles) are able to move with 
the fluid and their properties are distributed in space through a smoothing 
function that allows all hydrodynamic quantities to be continuous 
(\cite[Monaghan 1992]{monaghan92}). Our self-gravitating SPH code 
(\cite[Turner \etal\ 1995]{turner_etal95}) uses tree-code gravity 
(\cite[Barnes \& Hut 1986]{barnes_hut86}) to reduce the computational cost for 
the calculation of gravitational forces, and a second order Range-Kutta time 
integrator with multiple particle timesteps. It also uses Particle Splitting 
(\cite[Kitsionas \& Whitworth 2002]{kitsionas_whitworth02}), which allows the 
on-the-fly increase of numerical resolution (in terms of splitting the SPH 
particles) when and where this becomes necessary, i.e. every time that 
violation of the Jeans condition (\cite[Bate \& Burkert 1997]{bate_burkert97}, 
\cite[Truelove \etal\ 1997]{truelove_etal97}) becomes imminent. The benefit of 
the use of Particle Splitting is twofold. Firstly, gas fragmentation remains 
resolved at all times. Secondly, the simulations evolve with minimum 
computational cost, as high numerical resolution is employed only when 
required. Our Particle Splitting method assigns velocities to new particles by 
interpolating over the velocity field of their parent particle population, 
thus achieving conservation of energy and momenta with very high accuracy.

Our code 
includes sink particles (\cite[Bate, Bonnell \& Price 1995]{bate_etal95}). 
These are collisionless (star) particles that replace dense, collapsed gaseous 
regions in order to allow the simulations to evolve further in time. With sink 
particles, the simulations avoid using tiny timesteps to follow the evolution 
internal to the sink but at the same time all information on scales smaller 
than the size of the sink particle becomes invisible to us. In this work we 
use sink particles with radii $\sim 20$ AU. A summary of all features of our 
code can be found in 
\cite[Kitsionas \& Whitworth (2007)]{kitsionas_whitworth07}.

\section{Results}

\begin{figure}[t]
\begin{center}
\resizebox{6.5cm}{!}{\includegraphics{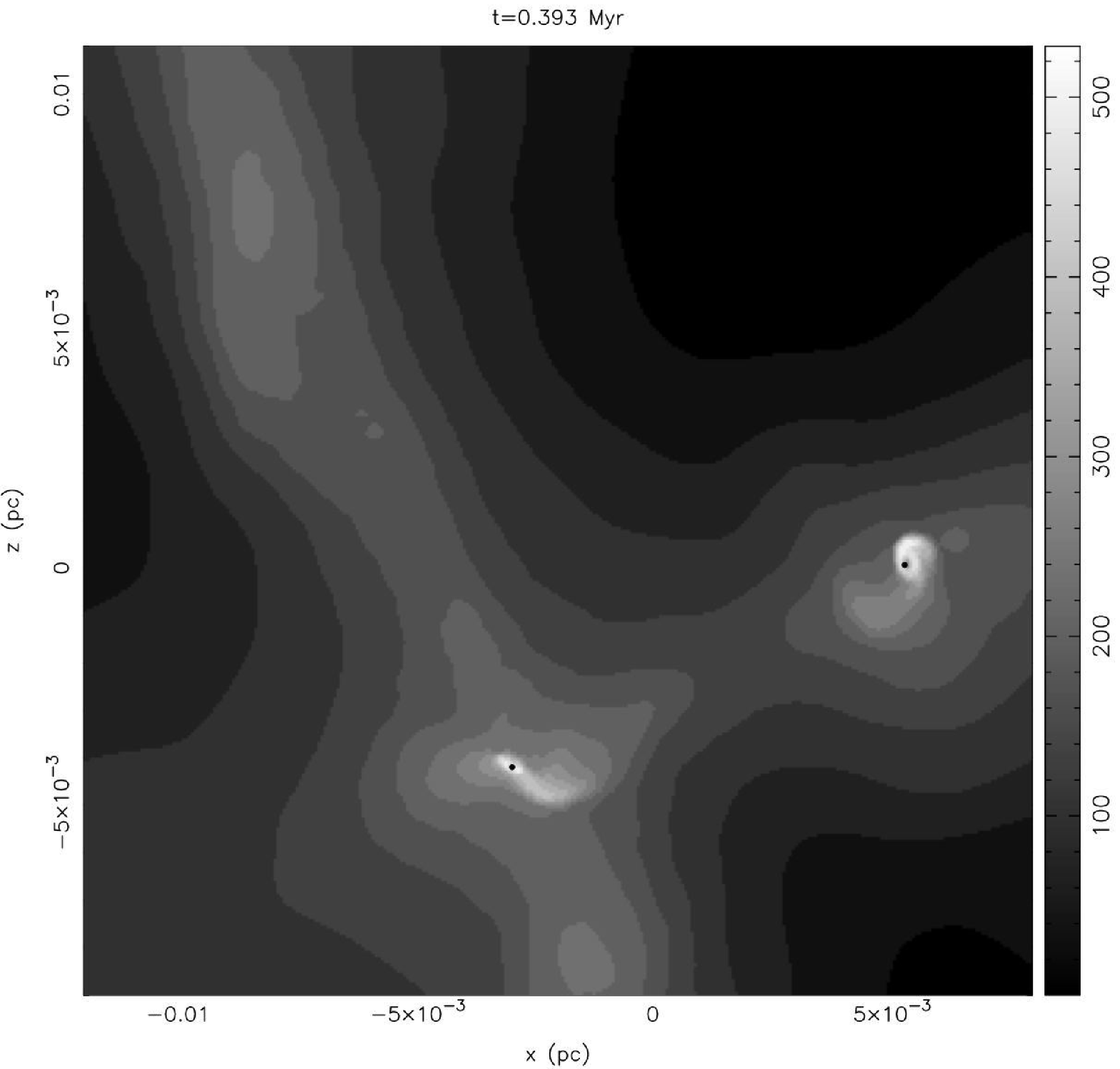} }
\resizebox{6.5cm}{!}{\includegraphics{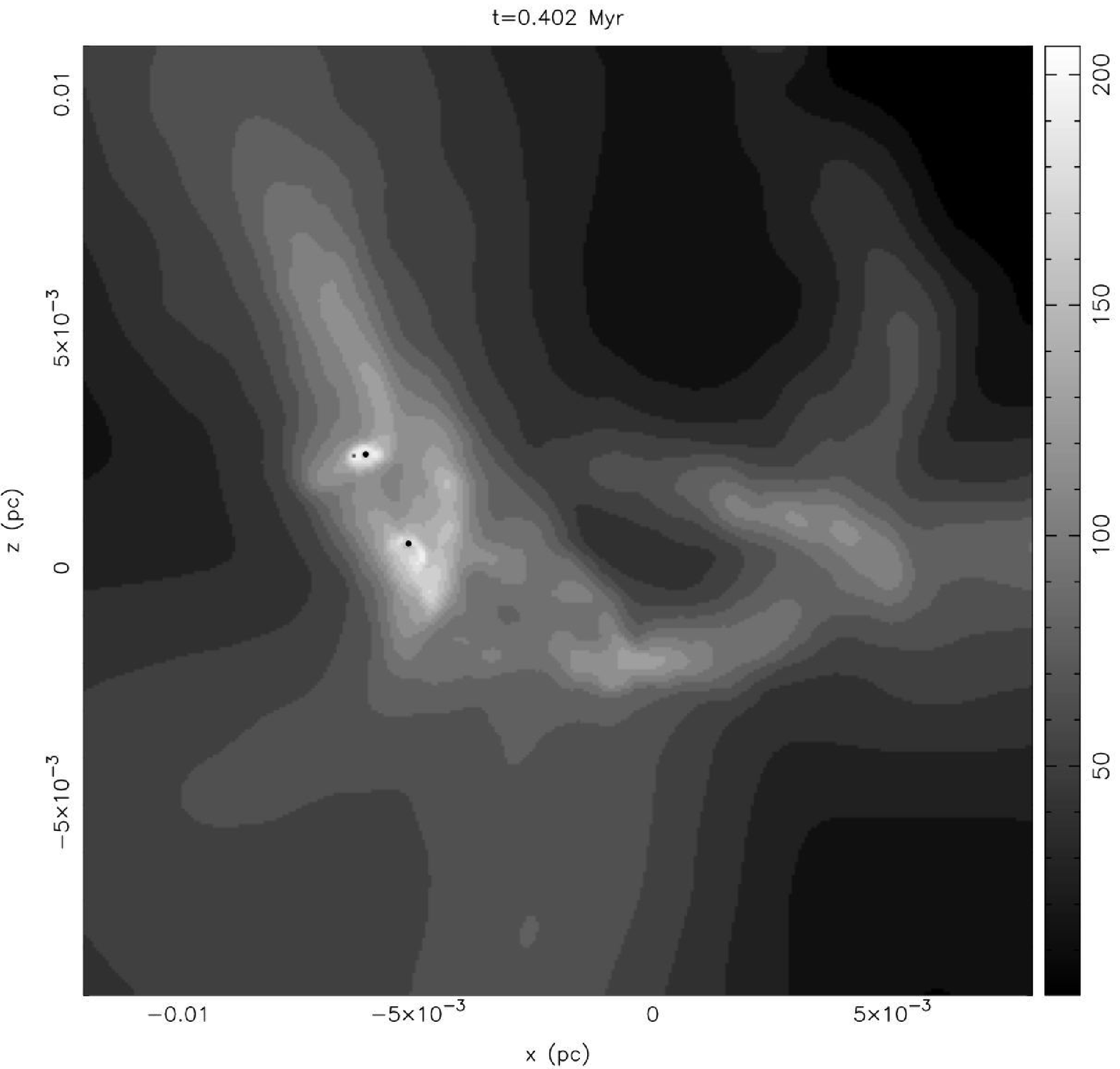} }
\caption{Column density plots for the collision with $M_0 = 10 M_\odot$, 
$b=0.2$, ${\cal M}=10$, with sink particles overlayed as solid circles 
(note that the symbol used for the sinks is larger than the actual sink radius 
at the scale of these plots). View along the $y$-axis; $\Delta x = \Delta z = 
0.02\,{\rm pc}$ in both panels. {\it Left Panel.} $t \sim 0.393\,{\rm Myr}$, 
i.e. very close to the time of the right panel of Fig. \ref{fig1}; 
sixteen-interval logarithmic grey-scale, in units of g cm$^{-2}$, from 
$3.31 \times 10^{-1} \, {\rm g} \, {\rm cm}^{-2}$ to 
$5.25 \times 10^{2} \, {\rm g} \, {\rm cm}^{-2}$. 
{\it Right Panel.} $t \sim 0.402\,{\rm Myr}$; 
sixteen-interval logarithmic grey-scale, in units of 
g cm$^{-2}$, from 
$2.88 \times 10^{-1} \, {\rm g} \, {\rm cm}^{-2}$ to 
$2.04 \times 10^{2} \, {\rm g} \, {\rm cm}^{-2}$.}
   \label{fig2}
\end{center}
\end{figure}

\begin{figure}[t]
\begin{center}
\resizebox{6.5cm}{!}{\includegraphics{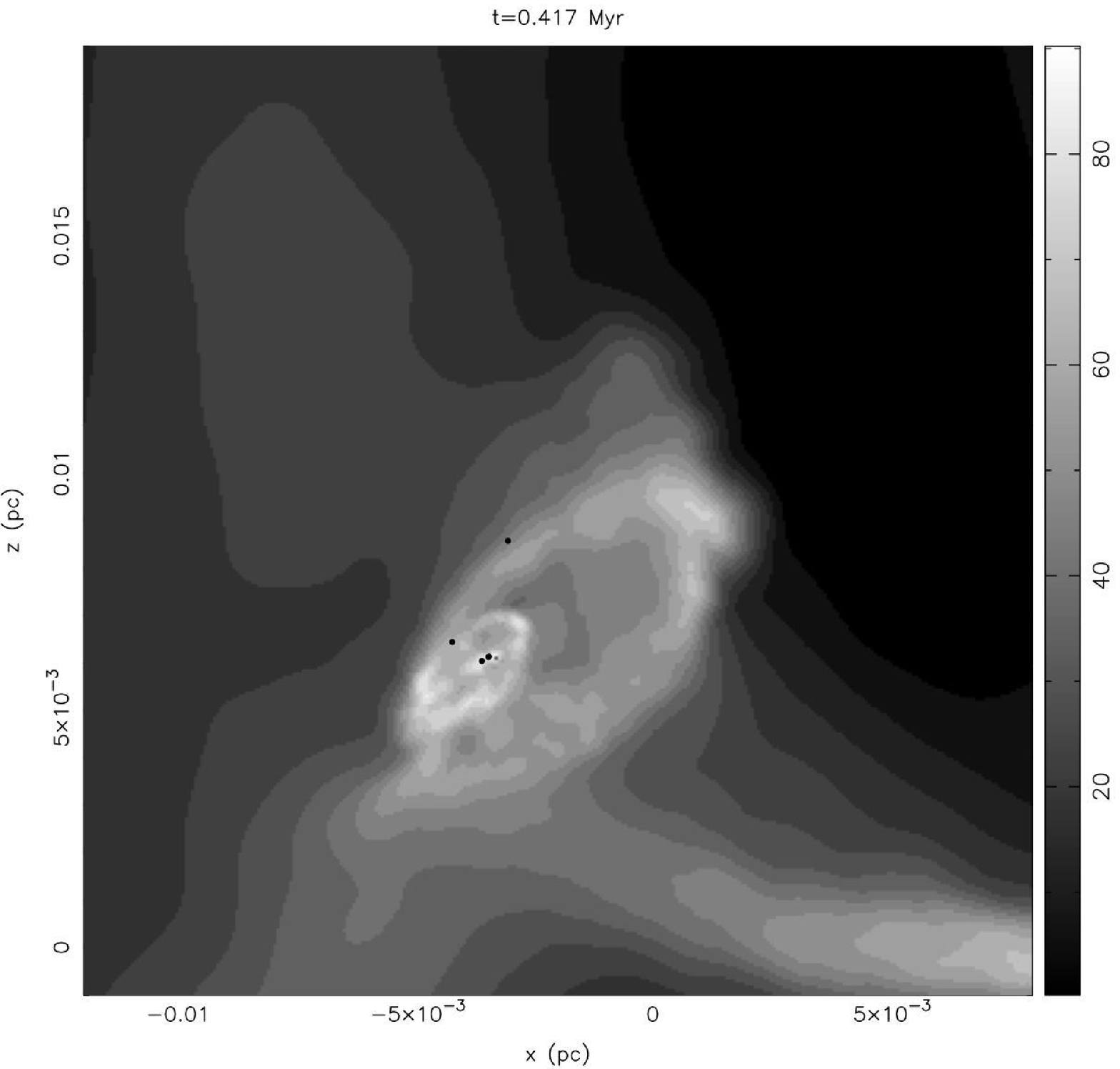} }
\resizebox{6.5cm}{!}{\includegraphics{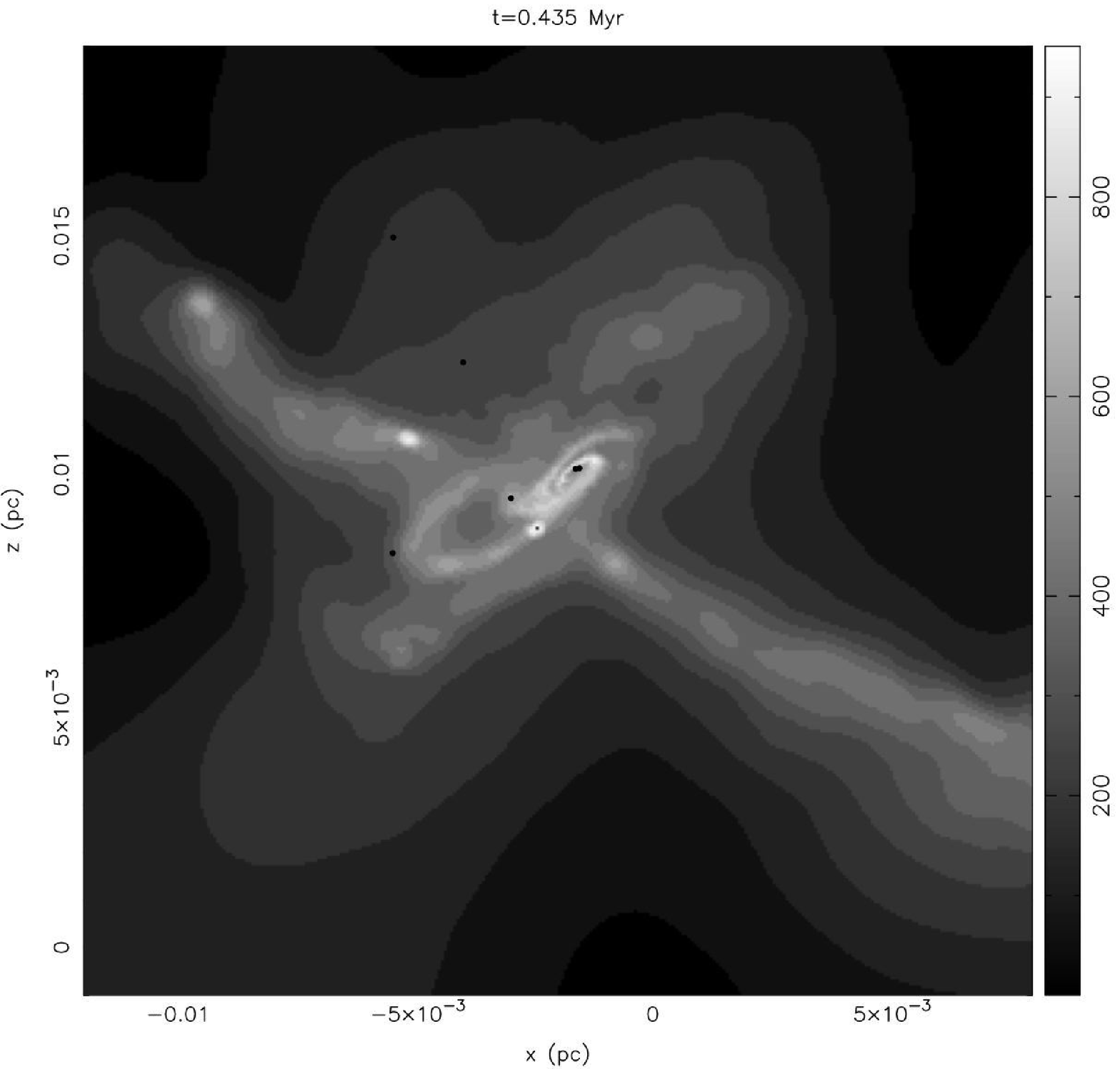} }
\caption{Column density plots for the collision with $M_0 = 10 M_\odot$, 
$b=0.2$, ${\cal M}=10$, with sink particles overlayed as solid circles 
(note that the symbol used for the sinks is larger than the actual sink radius 
at the scale of these plots). View along the $y$-axis; $\Delta x = \Delta z = 
0.02\,{\rm pc}$ in both panels (note that these panels have been shifted by 
0.008 pc along the $z$-axis with respect to the panels of Fig. \ref{fig2}). 
{\it Left Panel.} $t \sim 0.417\,{\rm Myr}$; 
sixteen-interval logarithmic grey-scale, in units of g cm$^{-2}$, from 
$2.45 \times 10^{-1} \, {\rm g} \, {\rm cm}^{-2}$ to 
$9.12 \times 10^{1} \, {\rm g} \, {\rm cm}^{-2}$. 
{\it Right Panel.} $t \sim 0.435\,{\rm Myr}$; 
sixteen-interval logarithmic grey-scale, in units of 
g cm$^{-2}$, from 
$3.09 \times 10^{-1} \, {\rm g} \, {\rm cm}^{-2}$ to 
$9.55 \times 10^{2} \, {\rm g} \, {\rm cm}^{-2}$.}
   \label{fig3}
\end{center}
\end{figure}

A shock-compressed layer forms at the collision interface as the collision 
proceeds. The right panel of Fig. \ref{fig1} offers a zoom-in view of this 
layer at $t \sim 0.396\,{\rm Myr}$. The layer extends along the $z$-axis. 
The $x$-$z$ projection of the shock-compressed layer at this time is shown in 
the left panel of Fig. \ref{fig2}. The layer has broken up into filaments and
the filaments have fragmented into protostars surrounded by circumstellar 
discs. Sink particles have replaced the central protostars of these discs. A 
second protostar forms on the upper part of the filament that extends almost 
vertically in this panel. At $t \sim 0.402\,{\rm Myr}$ (right panel of Fig. 
\ref{fig2}) the two protostars in this filament (represented by sink particles 
surrounded by 
circumstellar discs) have moved along the filament towards each other. 
Due to the tumbling nature of the filament, they have a close encounter in the 
course of 
which they capture each other into a binary. During this interaction, their 
discs merge to form a circumbinary disc (left panel of Fig. \ref{fig3}). This
disc-disc interaction leads also to the fragmentation of the circumbinary disc 
(see the additional sink particles in this panel). At later times, the 
circumbinary disc fragments {\it only} when lumps of material from the filament
fall on to the disc (e.g. see the lumps on the right panel of Fig. 
\ref{fig3}). 

\begin{figure}[t]
\begin{center}
\resizebox{6.5cm}{!}{\includegraphics{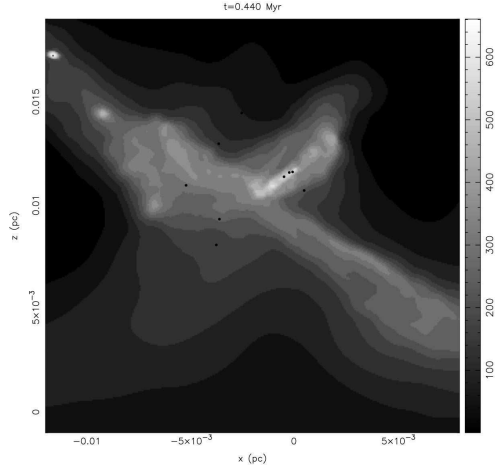} }
\caption{Column density plots for the collision with $M_0 = 10 M_\odot$, 
$b=0.2$, ${\cal M}=10$, with sink particles overlayed as solid circles 
(note that the symbol used for the sinks is larger than the actual sink radius 
at the scale of these plots). The final state of the simulation 
($t \sim 0.440\,{\rm Myr}$); view along the $y$-axis; $\Delta x = \Delta z = 
0.02\,{\rm pc}$; sixteen-interval logarithmic grey-scale, in units of 
g cm$^{-2}$, from $3.39 \times 10^{-1} \, {\rm g} \, {\rm cm}^{-2}$ to 
$6.61 \times 10^{2} \, {\rm g} \, {\rm cm}^{-2}$.} 
   \label{fig4}
\end{center}
\end{figure}

Each of the low-mass fragments, which form during the disc-disc interaction 
and/or during the episodes of lumpy accretion from the filament, gets ejected 
from the system after a close encounter with the binary. The final state of 
the system (Fig. \ref{fig4}) includes a large circumbinary disc hosting a 
binary, with components having mass $\sim 1.5 M_\odot$ (each) and separation 
$\lesssim 40$ AU, as well as a number 
of low-mass objects that have been ejected from the disc with moderate 
velocities of order a few km s$^{-1}$. Some of these low-mass objects are 
sub-stellar or even of planetary mass. After their ejection from the 
circumbinary disc they have stopped accreting more mass.

\section{Discussion}

Based on the results presented here, we can claim with confidence that 
dynamical interactions during the early stages of star formation provide a 
mechanism for the formation of low-mass objects. Disc-disc interactions and/or 
interactions between the discs and their gaseous environment (e.g. episodes 
of lumpy accretion from a filament) can lead to gravitational instabilities in 
protostellar discs and the formation of low-mass objects. Some of these 
objects may subsequently be ejected from the disc they formed in. As a result 
of this ejection, their accretion rates will drop significantly and they will 
end-up in the sub-stellar or even planetary-mass regime. 

It is important to note that the gas in our discs is rather warm, as no 
cooling mechanism for discs is included in our code and adiabatic gas heating 
has already switched on at the gas densities of our discs (see section 2 for a 
brief description of our equation of state). Therefore, the gravitational 
instabilities we report here are due to an increase in the disc surface 
density (caused by a dynamical interaction) and not because of the decrease 
of the gas sound speed in the disc. 

We note that the discs in our simulations contain only a few thousand SPH 
particles and evolve for a small number of orbital periods. Thus, due to 
excess numerical dissipation and limited orbital evolution, we are not yet in 
a position to investigate in detail the angular momentum evolution of gas and 
fragments in our discs. Nevertheless, fragmentation here is mostly externally 
triggered and therefore we are confident that it is not due to spurious 
angular momentum transport. Moreover, since the discs form self-consistently 
in a cluster environment, it is inevitable that during the initial stages of 
their formation they will contain a small but increasing number of particles. 
This makes impossible to comment on the effect of numerical dissipation 
on early disc evolution. Convergence studies with respect to the disc 
numerical dissipation in our simulations are currently underway.

Gas cooling is an important ingredient in simulations of isolated protostellar 
discs (e.g. see \cite[Rice \etal\ 2003]{rice_etal03}) even in the case that 
this cooling is not sufficient to produce gravitational instabilities and 
fragmentation of the disc (see the discussion in 
\cite[Durisen \etal\ 2007]{durisen_etal07}). 

\cite[Stamatellos, Hubber \& Whitworth (2007)]{stamatellos_etal07} have 
recently reported gravitational instabilities that lead to the formation of 
low-mass objects in an isolated disc using an SPH code with on-line 
(approximate) 
radiation transport. Their radiation hydrodynamics method is effective in 
calculating disc cooling close to the disc midplane but not further out in the 
disc ``atmosphere''. We are currently developing an improved algorithm that 
will be able to identify the scale-height at which each gas particle lies and 
thus assign the cooling efficiency of the particle accordingly.

Generalising our finding on the importance of dynamical interactions during 
the early stages of star formation, we assert that such interactions may be 
important for 
the formation of low-mass objects also at later stages of protostellar 
evolution, as long as a gas disc is still present. The outcome of such 
interactions need not be free-floating objects in all cases, as in less 
``chaotic'' situations (e.g. in circumstellar discs around single stars) the 
low-mass objects formed may remain bound to the system they formed in. 
Moreover, interactions of already formed planetary systems with incoming stars 
or lower mass free-floating objects may disturb the planetary systems, 
but this is beyond the scope of this paper (see several papers in this volume 
on planetary system dynamics).

Taking our results at face value, one can claim that the majority of the 
low-mass objects forming in such simulations will end up unbound from the 
system of their birth, i.e. a number of free-floating objects are expected to 
form as a by-product of star formation. 

The formation of free-floating planetary-mass objects is of particular 
interest. Such objects can be observed mainly through their gravitational 
signature. In the field, this signature could be detected only through 
microlensing observations. \cite[Zinnecker (2001)]{zinnecker01} has calculated 
that a lensing planet can produce a microlensing event with duration of a few 
days. He has also estimated that one such event is expected out of 
$3 \times 10^{8}$ observed lightcurves. He has further advocated that 
campaigns for the observation of free-floating planetary-mass objects through 
microlensing can become possible when the VST and VISTA facilities start their 
operation as well as with the JWST.

The event probability and duration that \cite[Zinnecker (2001)]{zinnecker01} 
has estimated are based on simple calculations involving the known frequency 
of binary stars with members hosting planetary systems and the probability for 
dynamical interactions between the members of such complicated systems. It 
would be interesting to calculate the corresponding numbers based on our 
models. However, we need conduct larger simulations of gravoturbulent star 
formation (see e.g. \cite[Klessen \& Burkert 2000, 2001]{klessen_burkert00}, 
\cite[Klessen 2001a,b]{klessen01a}, 
\cite[Jappsen \etal\ 2005]{jappsen_etla05}) 
in order to be able to derive more reliable statistics on the number of 
free-floating planetary-mass objects as well as the overall mass distribution 
of objects forming in the calculations. Such simulations are currently 
underway. Moreover, by using multiple levels of Particle Splitting we are now 
able to employ even higher numerical resolution for our discs as well as use 
smaller sink particles. We are thus in a position to investigate the evolution 
of individual discs within a realistic cluster environment at a resolution 
comparable to that of simulations of isolated discs. This way we can now 
address issues such as the amount of numerical dissipation and its effect on 
the angular momentum evolution of our discs.\\

\noindent {\bf Acknowledgments.} The authors would like to thank Hans 
Zinnecker for many interesting discussions. SK kindly acknowledges support by 
an EU Commission ``Marie Curie Intra-European (Individual) Fellowship'' of the 
6th Framework Programme. SK also acknowledges financial support by the German 
Research Foundation (DFG) for attending this symposium through travel grant 
KON 1577/2007.




\begin{thebibliography}{}

\bibitem[Barnes \& Hut (1986)]{barnes_hut86}
{Barnes, J., \& Hut, P.} 1986,
\textit{Nature}, 324, 446

\bibitem[Bate \& Bonnell (2005)]{bate_bonnel05}
{Bate, M.R., \& Bonnell, I.A.} 2005,
\textit{MNRAS}, 356, 1201

\bibitem[Bate \& Burkert (1997)]{bate_burkert97}
{Bate, M.R., \& Burkert, A.} 1997,
\textit{MNRAS}, 288, 1060

\bibitem[Bate, Bonnell \& Bromm (2003)]{bate_etal03}
{Bate, M.R., Bonnell, I.A., \& Bromm, V.} 2003,
\textit{MNRAS}, 339, 577

\bibitem[Bate, Bonnell \& Price (1995)]{bate_etal95}
{Bate, M.R., Bonnell, I.A., \& Price, N.M.} 1995,
\textit{MNRAS}, 277, 362

\bibitem[Bhattal \etal\ 1998]{bhattal_etal98}
{Bhattal, A.S., Francis, N., Watkins, S.J., \& Whitworth, A.P.} 1998,
\textit{MNRAS}, 297, 435

\bibitem[Bonnel \& Bate (2006)]{bonnel_bate06}
{Bonnell, I.A., \& Bate, M.R.} 2006,
\textit{MNRAS}, 370, 488

\bibitem[Chapman \etal\ (1992)]{chapman_etal92}
{Chapman, S.J., Pongracic, H., Disney, M.J., Neslon, A.H., Turner, J.A., \& Whitworth, A.P.} 1992,
\textit{Nature}, 359, 207

\bibitem[Durisen \etal\ (2007)]{durisen_etal07}
{Durisen, R.H., Boss, A.P., Mayer, L., Nelson, A.F., Quinn, T., \& Rice, W.K.M.} 2007, in: B. Reipurth, D. Jewitt, \& K. Keil (eds.),
\textit{Protostars and Planets V} (Tucson: University of Arizona Press), p. 607

\bibitem[Heitsch \etal\ (2005)]{heitsch_etal05}
{Heitsch, F., Burkert, A., Hartmann, L.W., Slyz, A.D., \& Devriendt, J.E.G.} 2005,
\textit{ApJ}, 633, L113

\bibitem[Hennebelle \& Audit (2007)]{hennebelle_audit07}
{Hennebelle, P., \& Audit, E.} 2007,
\textit{A\&A}, 465, 431

\bibitem[Hunter \etal\ (1986)]{hunter_etal06}
{Hunter, J.H. Jr., Sandford, M.T. II, Whitaker, R.W., \& Klein, R.I.} 1986,
\textit{ApJ}, 305, 309

\bibitem[Jappsen \& Klessen (2004)]{jappsen_klessen04}
{Jappsen, A.-K., \& Klessen, R.S.} 2004, 
\textit{A\&A}, 423, 1

\bibitem[Jappsen \etal\ (2005)]{jappsen_etal05} 
{Jappsen, A.-K., Klessen, R.S., Larson, R.B., Li, Y., \& Mac~Low, M.-M.} 2005, 
\textit{A\&A}, 435, 611

\bibitem[Kitsionas \& Whitworth (2002)]{kitsionas_whitworth02}
{Kitsionas, S., \& Whitworth, A.P.} 2002,
\textit{MMNRAS}, 330, 129

\bibitem[Kitsionas \& Whitworth (2007)]{kitsionas_whitworth07}
{Kitsionas, S., \& Whitworth, A.P.} 2007,
\textit{MMNRAS}, 378, 507

\bibitem[Klessen (2001a)]{klessen01a} 
{Klessen, R.S.} 2001a, 
\textit{ApJ}, 550, L77

\bibitem[Klessen (2001b)]{klessen01b}
{Klessen, R.S.} 2001b, 
\textit{ApJ}, 556, 837

\bibitem[Klessen \& Burkert (2000)]{klessen_burkert00}
{Klessen, R.S., \& Burkert, A.} 2000, 
\textit{ApJS}, 128, 287

\bibitem[Klessen \& Burkert (2001)]{klessen_burkert01}
{Klessen, R.S., \& Burkert, A.} 2001, 
\textit{ApJ}, 549, 386

\bibitem[Krumholz \& McKee (2005)]{krumholz_mckee05} 
{Krumholz, M.R., \& McKee, C.F.} 2005, 
\textit{ApJ}, 630, 250

\bibitem[Krumholz, McKee \& Klein (2005)]{krumholz_etal05}
{Krumholz, M.R., McKee, C.F., \& Klein, R.I.} 2005,
\textit{Nature}, 438, 332 [astro-ph/0510412]

\bibitem[Larson (1981)]{larson81}
{Larson, R.B.} 1981,
\textit{MNRAS}, 194, 809

\bibitem[Mac~Low \& Klessen (2004)]{maclow_klessen04} 
{Mac~Low, M.-M., \& Klessen, R.S.} 2004, 
\textit{Rev.\ Mod.\ Phys.}, 76, 125

\bibitem[Rice \etal\ (2003)]{rice_etal03}
{Rice, W.K.M., Armitage, P.J., Bate, M.R., \& Bonnell, I.A.} 2003,
\textit{MNRAS}, 338, 227 

\bibitem[Schmeja \& Klessen (2004)]{schmeja_klessen04}
{Schmeja, S., \& Klessen, R.S.} 2004,
\textit{A\&A}, 419, 405

\bibitem[Schmeja \& Klessen (2006)]{schmeja_klessen06}
{Schmeja, S., \& Klessen, R.S.} 2006,
\textit{A\&A}, 449, 151

\bibitem[Shu, Adams \& Lizano (1987)]{shu_etal87}
{Shu, F.H., Adams, F.C., \& Lizano, S.} 1987,
\textit{ARAA}, 25, 23

\bibitem[Stamatellos, Hubber \& Whitworth (2007)]{stamatellos_etal07}
{Stamatellos, D., Hubber, D.A., \& Whitworth, A.P.} 2007,
\textit{MNRAS}, 382, L30

\bibitem[Truelove \etal\ (1997)]{truelove_etal97}
{Truelove, K.J., Klein, R.I., McKee, C.F., Holliman, J.H., Howell, L.H., \& Greenough, J.A.} 1997,
\textit{ApJ}, 489, L179 

\bibitem[Turner \etal\ (1995)]{turner_etal95}
{Turner, J.A., Chapman, S.J., Bhattal, A.S., Disney, M.J., Pongracic, H., \& Whitworth, A.P.} 1995,
\textit{MNARS}, 277, 705

\bibitem[Vazquez-Semadeni \etal\ (2006)]{vazquez_etal06}
{V\'azquez-Semadeni, E., Ryu, D., Passot, T., Gonz\'alez, R.F., \& Gazol, A.} 2006,
\textit{ApJ}, 643, 245

\bibitem[Vazquez-Semadeni \etal\ (2007)]{vazquez_etal07}
{V\'azquez-Semadeni, E., G\'omez, G.C., Jappsen, A.-K., Ballesteros-Paredes, J., Gonz\'alez, R.F., \& Klessen, R.S.} 2007,
\textit{ApJ}, 657, 870

\bibitem[Walder \& Folini (1998)]{walder_folini98}
{Walder, R., \& Folini, D.} 1998,
\textit{A\&A}, 330, L21

\bibitem[Whitworth \etal\ (1995)]{whitworth_etal95}
{Whitworth, A.P., Chapman, S.J., Bhattal, A.S., Disney, M.J., Pongracic, H., \& Turner, J.A.} 1995,
\textit{MNARS}, 277, 727

\bibitem[Zinnecker (2001)]{zinnecker01}
{Zinnecker, H.} 2001, in: J.W. Menzies, \& P.D. Sackett (eds.),
\textit{Microlensing 2000: A New Era of Microlensing Astrophysics}, 
ASP Conference Proceedings, Vol. 239 (San Francisco: Astronomical Society of 
the Pacific), p. 223

\end{thebibliography}
\end{document}